\documentstyle[12pt]{article}
\newcommand{\beq}{\begin{equation}}
\newcommand{\eeq}{\end{equation}}
\newcommand{\beqa}{\begin{eqnarray}}
\newcommand{\eeqa}{\end{eqnarray}}
\newcommand{\beqar}{\begin{eqnarray*}}
\newcommand{\eeqar}{\end{eqnarray*}}

\newcommand{\norm}[1]{\raise.3ex\hbox{:}#1\raise.3ex\hbox{:}}

\renewcommand\d{{\rm d}}

\def\A{{\cal A}}
\def\B{{\cal B}}
\def\C{{\cal C}}

\renewcommand\d{{\rm d}}

\begin{document}
\begin{titlepage}
\rightline{\small IHES/P/97-03/24}
\vskip 5em

\begin{center}
{\bf \huge
Comments on Dirichlet Branes at Angles}
\vskip 3em

{\large Noureddine Hambli\footnote{email:
hambli@ihes.fr\hfil}}
\vskip 1em

{\em  Institut des Hautes Etudes Scientifiques\\
      35, route de Chartres, 91440 - Bures-sur-Yvette, France}	

\vskip 4em

\begin{abstract}
This paper illustrates the derivation of the low-energy background field
solutions of D2-branes and D4-branes intersecting at non-trivial angles by
solving {\it directly} the bosonic equations of motion of II supergravity
coupled to a dilaton and antisymmetric fields. We also argue for how a similar
analysis can be performed for any similar Dp-branes oriented at angles.
Finally, the calculation presented here serves as a basis in the search for a
systematic derivation of the background fields of the more general
configuration of a p-brane `angled' with a q-brane ($p \neq q$). 
\end{abstract}
\end{center}

\end{titlepage}

\setcounter{footnote}{0}
\section{Introduction}

Since the realization by Polchinski that they are 
the carriers of Ramond-Ramond (RR) charges~[1], 
D-branes contributed greatly to our understanding 
of the non-perturbative aspects of string theory.
For example, recently bound states of D-branes have 
proven to be useful in the statistical interpretation 
of the Bekenstein-Hawking entropy associated with 
certain supersymmetric black holes in string 
theory~[2]. In string theory, the large degeneracy of
a bound state can be counted reliably at weak coupling. 
Then as one increases the strength of the coupling, 
the bound state eventually undergoes a gravitational 
collapse and forms a black hole of string theory. 
In these analyses the bound states of interest must be
supresymmetric, {\it i.e.,} saturate a 
BPS bound, since the spectrum of supersymmetric 
states are protected against loop corrections as we 
vary the coupling. Therefore the degeneracy of the
black holes states at strong coupling must be the same
as that of the corresponding weakly coupled D-brane bound 
state\footnote{Recently in~[3], however, a similar counting 
of states was performed for a certain class of 
{\it non-supresymmetric} but extremal black holes. In this case,
supersymmetry alone does not protect the spectrum and yet
the degeneracy at strong coupling was shown to be the same as
that at weak coupling.}. This is one of the reasons that has 
made supersymmetric D-brane bound states a subject of growing 
interest.
\smallskip

Recently a number of BPS-saturated configurations representing
bound states of various Dp-brane solutions of 
type II string theory were obtained~[4]. In all these 
bound state solutions the constituent Dp-branes are either 
parallel to each other or intersect orthogonally according 
to the harmonic superposition rules where each brane is 
specified by an independent harmonic function~[5]. That these 
cannot be the only possibilities an that there exist
supresymmetric bound states where the D-branes intersect at 
angles which are different from zero and $\pi/2$ was first pointed 
out in~[6]. There it was shown that configurations of multiple 
D-branes related by $SU(N)$ rotations will preserve unbroken 
supersymmetry. The work of~[6] was later extended in~[7]
where it was demonstrated that there are toroidal 
compactifications of D-branes at angles in type II string theory 
that are supersymmetric only when non-trivial antisymmetric 
tensor moduli fields are turned on at the position of the 
D-brane. In both~[6] and~[7], the intervening configurations of 
D-branes at angles were analyzed using a string calculation. 
It is natural to try to demonstrate the findings of~[6] and~[7] 
from the point of view of the classical D-brane solution of supergravity. 
\smallskip

In effect, some research was conducted lately in which the 
classical solutions of branes oriented at angles appeared. 
We do not review any of these solutions here and refer the 
reader to the work in~[8,9]. For our present purpose, however, 
it suffices to notice that in the discussion of both~[8] and~[9], 
the authors started with a well known brane configuration and 
then used an appropriate combination of boosts, $T-$duality and 
$S-$duality transfomations to generate the supresymmetric bound 
state of D-branes at angles. In this paper, we will take 
a different approach and show how the low-energy background fields 
of D2-branes and D4-branes whose pairs are oriented at angles can be 
derived by solving {\it directly} the bosonic equations of motion 
of II supergravity coupled to a dilaton and antisymmetric fields.
\smallskip

The paper is organized as follows: We start by establishing
our conventions in Section~(2), as well as presenting the 
low-energy action for type IIA string theory and the equations
of motion derived from it. In Section~(3), by restricting to 
the case where only the R-R 4-form field strength is kept 
excited, we study in detail the configuration of two D2-branes 
by $SU(2)$ angles, and which preserves 1/4 of supersymmetry. 
In particular, we discuss the choice of the ans\"atze to be made 
for the background fields in order to solve the equations of motion. 
In Section~(4), we carry out the actual solving of the equations of 
motion and present all the calculational techniques that allow us 
to find the field solutions. We consider in Section~(5) 
the case of D4-branes where by also solving directly the 
equations of motion we find the low-energy solution of the 
corresponding bound state with angle and with 1/4 unbroken 
supresymmetry. We also argue how a similar analysis can be
performed for the general case of p-branes oriented at angles
in type II string theories and M-theory. Finally, a brief 
discussion follows in Section~(6).
\smallskip

\section{Some Preliminaries}

The bosonic part of the low-energy action for
type IIA string theory in ten dimensions is~[10]
\begin{eqnarray}
S_{IIA} & = & {1\over{16 \pi G_N}} \int d^{10}\!x \sqrt{-G}
\left[
R - {1\over 2} (\partial\phi)^2 -
{1\over2}\, \left({1\over{3!}}\, e^{-\phi}\, H^2 -
{1\over{2!}}\, e^{3 \phi /2}\, f^2 -
{1\over{4!}}\, e^{\phi /2}\, F^2\right)
\right] \nonumber\\
& &\qquad\qquad\qquad\qquad\qquad
- \; {1\over{32\, \pi\, G_N}}\, \int\, B\, dA\, dA\, , 
\label{aa}
\end{eqnarray}
where $G_{\mu\nu}$ is the Einstein-frame metric,
$H = dB$ is the 3-form field strength of the Kalb-Ramond
NS-NS field, $f = da$ and $F = dA - H a$ are the 2-form and
4-form R-R field strengths, and finally $\phi$ is the dilaton 
which is taken to vanish asymptotically. 
\smallskip

Dropping the various Chern-Simons terms
from $S_{IIA}$ (since the solutions we will present throughout
all the paper are also consistent solutions of the full action), and 
by setting both the field strength $H$ and $f$ to zero the field equations 
of motion take the simple form\footnote{Our conventions are 
$(-,+,\cdots,+)$ signature for the metric $G_{\mu\nu}$, for the
Riemann tensor we take
${{R^{\lambda}}_{\mu\,\nu}}_\kappa = 
\partial_\nu\, \Gamma^\lambda_{\mu\,\kappa} - 
\partial_\kappa\, \Gamma^\lambda_{\mu\,\nu} +
\Gamma^\eta_{\mu\,\kappa}\,\Gamma^\lambda_{\nu\,\eta} -
\Gamma^\eta_{\mu\,\nu}\,\Gamma^\lambda_{\kappa\,\eta}$
and for the affine connection we take
$\Gamma^\lambda_{\mu\,\nu} = G^{\lambda\,\rho} \,
\left(\partial_\mu\, G_{\rho\,\nu} +
\partial_\nu\, G_{\mu\,\rho} -
\partial_\rho\, G_{\mu\,\nu}
\right)/2$.}
\begin{equation}
{R^\mu}_\nu = {1\over 2}\, \partial^\mu \phi\, \partial_\nu \phi +
{1\over{2\, 4!}}\, e^{\phi/2}\,
\left[4\, F^{\mu\alpha_2\alpha_3\alpha_4}\,F_{\nu\alpha_2\alpha_3\alpha_4}\, -\,
{3\over 8}\, {G^\mu}_\nu \,  F^2\right]\, ,
\label{ab}
\end{equation}
\begin{equation}
\Box\phi = {1\over{4\, 4!}}\, e^{\phi/2}\, F^2\, ,
\label{ac}
\end{equation}
\begin{equation}
\partial_{\mu_1}\, 
\left(\sqrt{-G}\, e^{\phi/2}\, F^{\mu_1 \mu_2 \mu_3 \mu_4}
\right) = 0 \, .
\label{ad}
\end{equation}
The statement that the R-R 4-form $F$ is the field strength of a 
3-form potential $A$ is equivalent to the Bianchi identity imposed 
by the condition:
\begin{equation}
\partial_{[ \nu}\, F_{\mu_1 \mu_2 \mu_3 \mu_4 ]} = 0\, .
\label{ae}
\end{equation}
\smallskip

In general, in type II superstring theories there 
are two kind of p-branes, those charged under NS-NS fields and 
those carrying R-R charge. The first ones contain the usual
elementary (perturbative) string states in the case of 1-brane~[11] 
and the purely solitonic objects in the case of the 5-brane~[12]. 
For the second ones however, the branes have been shown 
to describe Dp-branes~[1]. Within type IIA theory,
the Dp-brane can have $p=0,2,4,6,8$, while for type IIB theory
p ranges over $-1,1,3,5,7,9$. Since only the R-R 4-form field 
strength $F$ is kept excited in our type IIA theory, then we are
certain that the low-energy background fields that solve the 
equations of motion derived from $S_{IIA}$ are those of D2-branes, 
{\it i.e.,} D-membranes. The D2-branes are {\it electrically} 
charged.
\smallskip
 
Our purpose in this paper is to find the low-energy background
fields describing two D2-branes intersecting at angles different 
from zero and $\pi/2$ by solving {\it directly} the equations of 
motion (\ref{ab},\ref{ac},\ref{ad},\ref{ae}). Recently using a 
similar strategy, a model-independent derivation for the general 
rules of orthogonnally intersecting extreme branes in arbitrary 
spacetime dimension~$D$ was given in~[13]. Following the same 
approach, the non-extreme case was also treated in~[14].
By specializing to the branes occuring in type II superstring 
theories and in M-theory, the intersection rules found in this way
are compatible with the harmonic superposition rules for intersecting 
extreme p-branes formulated originallly by Tseytlin in~[5]
and later generalized to non-extreme cases in~[14] 
(see also~[15]). 
\smallskip

\section{The Making of Two D2-Branes at Angles}

To obtain a brane configuration that correspond to two D2-branes 
intersecting at some non-trivial angle $\theta \ne 0, \pi/2$, 
we can start by two parallel D2-branes oriented both along 
the $x^{1,2}-$axes. Then we rotate one of them by the angle 
$\theta$ in the $x^{1,3}$ plane and $-\theta'$ in the $x^{2,4}$ 
palne using a rotation matrix which acts on the coordinates 
$x^{1,2,3,4}$ in the following way:
\begin{eqnarray}
y^1 & = & \cos \theta\; x^1 - \sin \theta \; x^3\; , \qquad
y^3  =  \sin \theta \; x^1 + \cos \theta \; x^3\, ,
\nonumber\\
y^2 & = &\cos \theta \; x^2 + \sin \theta \; x^4\; , \qquad
y^4  =  -\, \sin \theta \; x^2 + \cos \theta \; x^4 \; .
\label{af}
\end{eqnarray}
In the new coordinates $y^{1,2,3,4}$ defined above, 
one brane stays oriented along the $x^{1,2}-$axes while 
the other is now oriented along the $y^{1,2}-$axes. From 
the work of~[6], the new configuration obtained will
preserve some supresymmetry (1/4 in the present case)
if the relative orientation of the D2-branes are
restricted by an $SU(2)$ rotation. If one defines the 
complex coordinates $z^1 = x^1 + i x^3$ and $z^2 = x^2 + i x^4$, 
then the rotations in (\ref{af}) take the simple form
$\left( z^1 \, , \, z^2 \right) \to 
\left( e^{i\theta}\, z^1 \, , \,  e^{-i\theta'}\, z^2 \right)$.
It follows that if the D2-branes should be related by an $SU(2)$ 
rotation, {\it i.e.,} $z^i \to 
{{\left[ \exp \left( i\theta\, \sigma_3 \right) \right]}^i}_j \; z^j$,
then $\theta = \theta'$~[9]. From this construction it is clear that 
the general supergravity solution of D2-branes at angles is in effect
an interpolation between the parallel case ($\theta = 0$) and
the orthognal intersection case ($\theta = \pi/2$).
This observation will prove to be useful in section~(4) 
when we come to inetgrate the field equations of 
(\ref{ab},\ref{ac},\ref{ad},\ref{ae}). The reason for this is that
eventhough the equations of motion of are the simplest 
possible since they involve only the fields $\phi$, $G_{\mu\nu}$
and $F_{\mu_1\mu_2\mu_3\mu_4}$, it would not be possible
to solve them directly without having recourse to make few ans\"atze. 
To build our intuition on the kind of ans\"atze that one should 
be making, we review below the supergravity solutions 
of two parallel D2-branes, two D2-branes intersecting orthogonally
over a point and a single rotated D2-brane.
\smallskip

\subsection{Parallel D2-Branes}

In the Einstein-frame metric, the low-energy background field 
solution~[16,17] describing two parallel D2-branes oriented along
$x^{1,2}-$axes contains only a non-trivial metric, dilaton 
and a single R-R potential, $A$ as follows:
\begin{eqnarray}
\d s^2 & = &\, \left(1 + {h_1} + {h_2}\right)^{{3}\over 8}\,
\left({{- \d t^2 + \left( \d x^1 \right)^2 + 
\left( \d x^2 \right)^2}\over{1 + {h_1} + {h_2}}}
+ \left(\d x^3 \right)^2 + \left(\d x^4 \right)^2 + \cdots + 
\left(\d x^9 \right)^2 \right)\, , \nonumber\\
A & = &\, {{{h_1} + {h_2}}\over{1 + {h_1} + {h_2}}} 
\d t\wedge\d x^1\wedge \d x^2 \, ,\nonumber\\
F_{t12i} & = &\, -\, \left({1\over{1 + h_1 + h_2}}\right)^2\, \partial_i h_1
-\, \left({1\over{1 + h_1 + h_2}}\right)^2\, \partial_i h_2\, ,
\nonumber\\
e^{2\phi} & = &\, \sqrt{1 + {h_1} + {h_2} }\, .
\label{ag}
\end{eqnarray}
The spatial coordinates $\left(x^1 , x^2\right)$ are parallel 
to the worldvolume of the D2-branes, while the orthogonal space 
is spanned by the coordinates $x^i$ with $i=3,4,...,9$. 
A noteworthy feature of the above solution is that it is 
completely specified by the harmonic functions ${h_{1,2}}$ 
which solve the flat space Poisson's equation in 
the transverse space:
\begin{equation}
\delta^{ij}\, \partial_i \, \partial_j \, h_{1,2} = 0\, .
\label{ah}
\end{equation}
The solutions of (\ref{ah}) has a dependence only through
the radius $r^2 = \displaystyle{\sum_{i=3}^{9}\, (x^i)^2}$ 
in the transverse space and is given by:
\begin{equation} 
{h_{1,2}} \equiv {h_{1,2}}(r) = 
{{c_{1,2} \, Q_{1,2}}\over{\left| \vec{r} - \vec{r}_{1,2} \right|^5}} \, ,
\label{ai}
\end{equation}
where $\vec{r}_1$ is the location in the transverse space of
the first D2-brane and $\vec{r}_2$ is the location of the second.
\smallskip

\subsection{Orthognal D2-Branes}

In the Einstein-frame metric, the low-energy background field 
solution [5,13] describing two orthogonal D2-branes 
intersecting over a point, where one of them is oriented along 
$x^{1,2}-$axes and the other one oriented along $x^{3,4}-$axes 
is given by:
\begin{eqnarray}
\d s^2 & = &\, \left(1 + {h_1} + {h_2} + {h_1}\,{h_2}\right)^{{3}\over 8}\,
\left( {{- \d t^2}\over{1 + {h_1} + {h_2} + {h_1}\,{h_2}}} + 
{{\left( \d x^1 \right)^2 + \left( \d x^2 \right)^2}\over{1 + {h_1}}}
\right.
\nonumber\\
& &\quad\quad\;\; \left. + {{\left(\d x^3 \right)^2 + 
\left(\d x^4 \right)^2}\over{1 + {h_2}}} 
+ \left(\d x^5 \right)^2 + \cdots + \left(\d x^9 \right)^2 \right)\, , 
\nonumber\\
A & = &\, {{h_1}\over{1 + {h_1} }}\,
\d t\wedge\d x^1\wedge \d x^2 \, -\, 
{{h_2}\over{1 + {h_2} }}\,
\d t\wedge\d x^3\wedge \d x^4 \, \nonumber\\
F_{t12i} & = &\, -\, \left({1\over{1 + h_1}}\right)^2\, \partial_i h_1\, ,
\nonumber\\
F_{t34i} & = &\,\left({1\over{1 + h_2}}\right)^2\, \partial_i h_2\, ,
\nonumber\\
e^{2\phi} & = &\, \sqrt{1 + {h_1} + {h_2} + {h_1}\,{h_2} }\, .
\label{aj}
\end{eqnarray}
For this solution, the harmonic functions $h_{1,2}$ 
satisy the Poisson's equation in the transverse subspace 
spanned by $\left( x^5 , x^6 , x^7 , x^8 , x^9 \right)$:
\begin{equation}
\delta^{ij}\, \partial_i\, \partial_j\,
h_{1,2} = 0\, \qquad \hbox{for}\quad i = 5, \cdots , 9\, ,
\label{ak}
\end{equation}
yielding the solutions
\begin{equation} 
{h_{1,2}} \equiv {h_{1,2}}(r) = 
{{c_{1,2} \, Q_{1,2}}\over{\left| \vec{r} - \vec{r}_{1,2} \right|^3}} \, ,
\label{al}
\end{equation}
where here $r^2 = \displaystyle{\sum_{i=5}^{9}\, (x^i)^2}$. 
\smallskip

\subsection{A Single Rotated D2-Brane}

In (\ref{ag}) or (\ref{aj}), if we set $h_2 = 0$ we recover 
the background field solution of a single D2-Brane oriented 
along the $x^{1,2}-$axes. Now if we perform a rotation to orient 
this D2-brane along the $y^{1,2}-$axes which were introduced in (\ref{af}), 
then the new configuration is given by:
\begin{eqnarray}
\d s^2 & = &\, \left(1 + {h_1}\right)^{3\over 8}\,
\left({{- \d t^2}\over{1 + {h_1}}} + 
{{1 + {h_1} \sin^2 \theta}\over{1 + {h_1}}}\, 
\left[
\left(\d x^1\right)^2 + \left(\d x^2\right)^2
\right] \right.
\nonumber\\
& &\;\; \left.+ {{1 + {h_1} \cos^2 \theta}\over{1 + {h_1}}}\, 
\left[
\left(\d x^3\right)^2 + \left(\d x^4\right)^2
\right] + 2\, \cos \theta\, \sin \theta\, 
{{h_1}\over{1 + {h_1}}}\,
\left(
\d x^1\, \d x^3 - \d x^2\, \d x^4
\right) \right.
\nonumber\\
& & \qquad\qquad\qquad\qquad\qquad\qquad\qquad \qquad
\left. + \left(\d x^5 \right)^2 + \cdots + \left(\d x^9 \right)^2 \right)\, ,
\nonumber\\
A & = & 
{{{h_1}\, \cos^2 \theta}\over{1 + {h_1}}}\,
\d t \wedge \d x^1 \wedge \d x^2 +
{{{h_1}\, \cos \theta\, \sin \theta}\over{1 + {h_1}}}\,
\left(
\d t \wedge \d x^2 \wedge \d x^3 + \d t \wedge \d x^1 \wedge \d x^4
\right) 
\nonumber\\
& &\; 
- \; {{{h_1}\, \sin^2 \theta}\over{1 + {h_1}}}\,
\d t \wedge \d x^3 \wedge \d x^4\, ,
\nonumber\\
e^{2\phi} & = & \sqrt{1 + {h_1}} \, .
\label{am}
\end{eqnarray}
The D2-brane above is originally oriented along the $x^{1,2}-$axes, 
and normally, we would choose the harmonic functions $h_{1,2}$ to be on 
the whole tranverse space $\left( x^3 , x^4 , \cdots , x^9 \right)$ 
as in (\ref{ai}). For the present solution (\ref{am}), however, 
since the D2-brane is delocalized in the $x^{3,4}$ coordinates
the harmonic functions $h_{1,2}$ are given rather by (\ref{al}).
\smallskip

Next we display the field strengh 
which couples to this `tilted' D2-brane. 
A straightforward calculation using the R-R 3-form potential 
$A$ given in (\ref{am}) yields:
\begin{eqnarray}
F_{t12i} & = & -\; \left({{\cos \theta}\over{1 + {h_1}}}\right)^2 \;
\partial_i \, {h_1} \equiv -\, \left({{\A}\over{1 + h_1}}\right)^2 \, 
\partial_i \, {h_1}\, ,
\nonumber\\
F_{t23i} & = & F_{t14i} = -\; \left({{\cos \theta}\over{1 + {h_1}}}\right)\;
\left({{\sin \theta}\over{1 + {h_1}}}\right) \;
\partial_i \, {h_1} \equiv -\, \left({{\C}\over{1 + h_1}}\right)^2 \, 
\partial_i \, {h_1} \, ,
\nonumber\\
F_{t34i} & = & \left({{\sin \theta}\over{1 + {h_1}}}\right)^2 \;
\partial_i \, {h_1} \equiv \left({{\B}\over{1 + h_1}}\right)^2 \, 
\partial_i \, {h_1} \, ,
\label{an}
\end{eqnarray}
where $\partial_i \, {h_1} = \partial {h_1}/\partial {x^i}$,
for $i = 5 , \cdots , 9$. A glance at the expressions of the 
field strength above and the expression of the metric in
(\ref{am}) reveals the following relations: 
  \begin{eqnarray}
\C^2 & = & \A\, \B \, ,
\label{ao}\\
G_{11} & = & G_{22} = {{1 + {h_1}\, \sin^2 \theta}\over{1 + {h_1}}}\, ,
\label{ap}\\
G_{33} & = & G_{44} = {{1 + {h_1}\, \cos^2 \theta}\over{1 + {h_1}}}\, ,
\label{aq}\\
G_{13} & = & -\, G_{24} = A_{t23} = A_{t14} = 
\cos \theta\, \sin \theta\; {{h_1}\over{1 + {h_1}}}\, .
\label{ar}
\end{eqnarray}
These relations will prove their worth below when we come
to discuss the choice of the necessary ans\"atze involved
in the resolution of the equations of motion to find the 
field solution of the two D2-branes at angles.
\smallskip

\subsection{The Choice of the Ans\"atze}

To choose an ans\"atz for the background fields
$\left( G_{\mu\nu} , F_{\mu_1 \mu_2 \mu_3 \mu_4} , \phi \right)$ 
of the bound state of two D2-branes at angles, we refer back to 
the construction of this configuration proposed at the 
beginning of Section~(3). Following this construction, we
take one of the D2-branes, which we represent by the harmonic 
function $h_1$, to be oriented along the $y^{1,2}-$axes and the 
other one, which we represent by the harmonic function $h_2$, is 
set parallel to the $x^{1,2}-$axes. (The $y^{1,2}$ coordinates
are the same as defined in (\ref{af}).) This way of
representing the D2-branes at angles
has the advantage of yielding the well known
supregravity solution of Section~(3.3) when $h_2 = 0$. 
Based on these simple remarks and taking advantage of the relations of 
(\ref{ap},\ref{aq},\ref{ar}), we suggest the following 
starting ans\"atz for the background fields 
of two `angled' D2-branes expressed in the Einstein-frame 
metric:
\begin{eqnarray}
\d s^2 & = & -\, U^2 \d t^2 
+ G_{11}\, 
\left( \left(\d x^1\right)^2 +  \left(\d x^2\right)^2 \right)
+ G_{33}\,
\left( \left(\d x^3\right)^2 +  \left(\d x^4\right)^2 \right)
\nonumber\\
& & \qquad\qquad + 2\, G_{13}\, 
\left( \d x^1\, \d x^3 - \d x^2\, \d x^4 \right)
+ V^2\, \left(\d x^i\, \d x_i\right)\, ,
\nonumber\\
F_{t12i} & = & -\, \left({{\A_1}\over{E}}\right)^2 \, \partial_i h_1 
-\, \left({{\A_2}\over{E}}\right)^2 \, \partial_i h_2 \, ,  
\nonumber\\
F_{t23i} & = & F_{t14i} = 
-\, \left({{\C_1}\over{E}}\right)^2 \, \partial_i h_1 
+\, \left({{\C_2}\over{E}}\right)^2 \, \partial_i h_2 \, ,  
\nonumber\\
F_{t34i} & = & \left({{\B_1}\over{E}}\right)^2 \, \partial_i h_1 
+ \left({{\B_2}\over{E}}\right)^2 \, \partial_i h_2 \, ,  
\nonumber\\
e^{2\phi} & = & \sqrt{E}\, ,
\label{as}
\end{eqnarray}
where 
\begin{eqnarray}
U & \equiv & U\left(h_1 , h_2 , \theta\right),\;\;
V \equiv V\left(h_1 , h_2 , \theta\right),\;\;
E \equiv E\left(h_1 , h_2 , \theta\right), 
\nonumber\\
G_{11} & \equiv & G_{11}\left(h_1 , h_2 , \theta \right),\;\;
G_{13} \equiv G_{13}\left(h_1 , h_2 , \theta \right),\;\;
G_{33} \equiv G_{33}\left(h_1 , h_2 , \theta \right),
\nonumber\\
\A_m & \equiv & \A_m\left(h_1 , h_2 , \theta\right),\;\;
\B_m \equiv \B_m\left(h_1 , h_2 , \theta\right), \;\;
\C_m \equiv \C_m\left(h_1 , h_2 , \theta\right),\;\; m = 1,2\; ,
\label{at}
\end{eqnarray}
and the harmonic functions are 
${h_{1,2}} = {{c_{1,2} \, Q_{1,2}}/{\left| \vec{r} - \vec{r}_{1,2} \right|^3}}$,
with $r^2 = \displaystyle{\sum_{i=5}^{9}\, (x^i)^2}$.
\smallskip

All the unknown functions entering in the definition
of the fields above are represented by the set
$\left(U, V, G_{11} , G_{13} , G_{33} , \A_1 , \A_2 ,
\B_1 , \B_2 ,\C_1 , \C_2 , \phi\right)$ and our task in what 
follows will be to solve for them using the equations of motion. 
As we shall now discuss, the number of unknowns in this set can 
be reduced further by making further ans\"atze. First 
of all, let us note that (by construction) if we set 
$\theta = 0$ our bound state must reduce to that of Section~(3.1) 
where both the D2-branes are lying in the $\left(x^1 , x^2\right)$ 
plane and delocalized in the ${x^{3,4}}-$directions. Similarly, 
the configuration of two orthogonally D2-branes of Section~(3.2) 
is reproduced by choosing $\theta = \pi/2$. Finally, the special 
case $h_2 = 0$ corresponds to the `tilted' D2-brane of 
Section~(3.3). Because of these correspondences, we define our first 
simplifying ans\"atze, which accompany the low-energy solution above, 
by imposing the following constraints:
\begin{equation}
\left(\C_1\right)^2 = \A_1\, \B_1 \, ,\quad\quad
\left(\C_2\right)^2 = \A_2\, \B_2 \; ,
\label{au}
\end{equation}
\begin{equation}
\partial_1^2\, E =  \partial_2^2\, E = 0\; , 
\label{auu}
\end{equation}
where $\partial_{1,2} E \equiv \partial E/\partial h_{1,2}$.
Thus, we can now drop from our list of unknowns the functions 
$\C_1$ and $\C_2$. 
\smallskip

The third ans\"atz that we will make 
to reduce even further the number of unknowns is related 
to the condition of extremality. The requirement that the 
configuration of the D2-branes at angles preserves 1/4 of 
supresymmetries translates into the
following condition on the metric components:
\begin{equation}
U\; \left(G_{11}\, G_{33} - G_{13}^2\right)\; V^3 = 1\, .
\label{av}
\end{equation}
This equation is a generalization of the extremality condition
used in~[13] for the case of a bound state of two orthogonal 
D-branes. With this ans\"atz our list of unknown functions
has reducedto $\left(U, G_{11} , G_{13} , G_{33} , \A_1 , \A_2 ,
\B_1 , \B_2 , \phi\right)$. We are now ready to start solving the 
equations of motion using the ans\"atze introduced in this section.
\smallskip

\section{Solving the Equations of Motion}

Using the ans\"atz solution of (\ref{as}) accompanied by the
condition in (\ref{au}), the equations of motion of 
(\ref{ab},\ref{ac},\ref{ad}) simplify considerably and become:
\begin{equation} 
{1\over {4!}}\, e^{\phi/2}\, F^2 = 
-\, V^{-2} \, \left(S_{11}\, \left({\partial_i\, h_1}\right)^2 +
S_{22}\, \left({\partial_i\, h_2}\right)^2 +
2\, S_{12}\, \left({\partial_i\, h_1}\right)\left({\partial_i\, 
h_2}\right)\right)\, ,
\label{ba}
\end{equation}
\begin{eqnarray}
\partial_i \partial_i \ln\, U & = & {5\over{16}}\left[
S_{11} \left({\partial_i h_1}\right)^2 +
S_{22} \left({\partial_i h_2}\right)^2 +
2 S_{12} \left({\partial_i h_1}\right) \left({\partial_i h_2}\right)\right],  
\label{bb}\\
\partial_i \partial_i  \ln \left( G_{11} G_{33} - G_{13}^2 \right) & = & 
{1\over{4}}
\left[ 
S_{11} \left({\partial_i h_1}\right)^2 +
S_{22} \left({\partial_i h_2}\right)^2 +
2 S_{12} \left({\partial_i h_1}\right)\left({\partial_i h_2}\right)\right],
\label{bc}\\
\partial_i \partial_i \phi & = & -\, {1\over{4}}
\left[ 
S_{11} \left({\partial_i h_1}\right)^2 +
S_{22} \left({\partial_i h_2}\right)^2 +
2 S_{12} \left({\partial_i h_1}\right) \left({\partial_i h_2}\right)\right],
\label{bd}
\end{eqnarray}
\begin{eqnarray}
& & \partial_i \ln U\,\partial_j \ln U  -  
{1\over 2} \left(\partial_i G^{11}\, \partial_j G_{11} + \partial_i G^{33}\, 
\partial_j G_{33}
+ 2 \partial_i G^{13}\, \partial_j G_{13}\right) +
3 \partial_i \ln V\,\partial_j \ln V
\nonumber\\
& &\;\; + \delta_{ij}\, \partial_k \partial_k \ln V = 
- {1\over 2} \partial_i \phi\, \partial_j \phi -\, {3\over{16}} \delta_{ij}
\left[S_{11} \left(\partial_k h_1\right)^2 + S_{22} \left(\partial_k 
h_2\right)^2
+ 2 S_{12} \partial_k h_1 \partial_k h_2 \right]
\nonumber\\
&&\qquad\quad\quad \qquad  
+ {1\over 2} \left[S_{11} \partial_i h_1 \partial_j h_1 + S_{22} \partial_i h_2 
\partial_j h_2
+ S_{12} \partial_i h_1 \partial_j h_2 + S_{12} \partial_i h_2 \partial_j 
h_1\right]\, ,
\label{bcc}
\end{eqnarray}
with
\begin{eqnarray}
S_{12} & = & 
{ { \left(E^{31/16} \, U\right)^{-2} } \over {\left(G_{11}\, G_{33} - 
G_{13}^2\right)^2} }\,  
\left[G_{33}\, \A_1\, \A_2 - G_{11}\, \B_1\, \B_2 - G_{13}\left(\A_1\, \B_2 - 
\A_2\, \B_1\right)\right]^2 \, ,
\label{be}\\
S_{11} & = & 
{{\left(E^{31/16}\,U\right)^{-2}}\over{\left(G_{11}\, G_{33} - 
G_{13}^2\right)^2}}\,  
\left[G_{33}\, \A_1^2 + G_{11}\, \B_1^2 + 2\, G_{13}\, \A_1\,\B_1\right]^2 \, ,
\label{bf}\\
S_{22} & = & 
{{\left(E^{31/16}\,U\right)^{-2}}\over{\left(G_{11}\, G_{33} - 
G_{13}^2\right)^2}}\, 
\left[G_{33}\, \A_2^2 + G_{11}\, \B_2^2 - 2\, G_{13}\, \A_2\,\B_2\right]^2 \, .
\label{bg}
\end{eqnarray}
As an immediate consequence of the equations (\ref{bb},\ref{bc},\ref{bd}) and 
the 
extremality condition (\ref{av}), we have 
\begin{equation}
U = E^{-\, 5/16}\, ,\quad V = E^{3/16}\, , \quad 
\left(G_{11}\, G_{33} - G_{13}^2\right) = E^{-\, 1/4}\, .
\label{bh}
\end{equation}
These relations simplify furher our problem and allows us to keep only 
the set of functions $\left(G_{11} , G_{13} , G_{33} , \A_1 , \A_2 , \B_1 , \B_2 
\right)$. 
in our list of unknowns.
\smallskip

Since the dilaton $\phi$ has been dropped from our list of unknowns 
we will in a first step determine the expressions of the metric 
components $G_{11}$, $G_{13}$ and $G_{33}$ in terms of it. For this, 
we begin by showing using the harmonic property of the functions 
$h_{1,2}$, {\it i.e.,} $\partial_i\, \partial_i\, h_{1,2} = 0$, and 
the relation $4\,\phi = \ln {E}$ from the ans\"atz in (\ref{as}) that:
\begin{equation}
4\, \partial_i \partial_i  \phi = -\left({{\partial_1 E}\over E}\right)^2 
\left(\partial_i h_1\right)^2
-\left({{\partial_2  E}\over E}\right)^2 
\left(\partial_i h_2\right)^2 
- 2 {{\partial_1 E \partial_2 E - E \partial_1\partial_2 E}\over{E^2}}
\left({\partial_i h_1}\right) \left({\partial_i h_2}\right) ,
\label{bi}
\end{equation}
where we used the ans\"atz (\ref{auu}) to set 
$\partial_1^2 E = \partial_2^2 E = 0$. Comparing equation
(\ref{bi}) with the equation of motion of $\phi$ in (\ref{bd}) 
and using the relations (\ref{be},\ref{bf},\ref{bg}) for $S_{11}$, 
$S_{13}$ and $S_{33}$ after substituting also by the expressions in 
(\ref{bh}), we obtain: 
\begin{eqnarray}
E^{3/4} \left(\partial_1 E \partial_2 E - E \partial_1\partial_2 E\right) & = &   
\left[G_{33} \A_1 \A_2 - G_{11} \B_1 \B_2 - G_{13}\left(\A_1 \B_2 - 
\A_2 \B_1\right)\right]^2  ,
\label{bj}\\
E^{3/4}\, \left(\partial_1 E\right)^2 & = &   
\left[G_{33}\, \A_1^2 + G_{11}\, \B_1^2 + 2\, G_{13}\, \A_1\,\B_1\right]^2 \, ,
\label{bk}\\
E^{3/4}\, \left(\partial_2 E\right)^2 & = & 
\left[G_{33}\, \A_2^2 + G_{11}\, \B_2^2 - 2\, G_{13}\, \A_2\,\B_2\right]^2 \, .
\label{bl}
\end{eqnarray}
Now if we take the square root of the three equations above we are led 
remarkably to a simple system of three simultaneous linear equations 
where we can at last solve for $G_{11}$, $G_{13}$ and $G_{33}$
in terms of $\left(\A_1 , \A_2 , \B_1 , \B_2 , E\right)$. Thus, we have
\begin{eqnarray}
E^{3/8} \sqrt{\partial_1 E \partial_2 E - 
E \partial_1\partial_2 E} & = &   
G_{33} \A_1 \A_2 - G_{11} \B_1 \B_2 - G_{13}\left(\A_1 \B_2 - 
\A_2 \B_1\right)\,  ,
\label{bm}\\
E^{3/8}\, \partial_1 E & = &   
G_{33}\, \A_1^2 + G_{11}\, \B_1^2 + 2\, G_{13}\, \A_1\,\B_1\, ,
\label{bn}\\
E^{3/8}\, \partial_2 E & = & 
G_{33}\, \A_2^2 + G_{11}\, \B_2^2 - 2\, G_{13}\, \A_2\,\B_2\, .
\label{bo}
\end{eqnarray}
Solving for the metric components we find:
\begin{eqnarray}
G_{13} & = &
{ E^{3/8} {\left[
\left(\partial_1 E\right) \A_2 \B_2 - 
\left(\partial_2 E\right) \A_1 \B_1 -
\sqrt{\partial_1 E \partial_2 E - E \partial_1\partial_2 E}\; 
\left(\A_1 \B_2 - \A_2 \B_1\right)
\right]}\over
{\left( \A_1 \B_2 + \A_2 \B_1 \right)^2}}\, ,
\nonumber\\
G_{11} & = &
{ E^{3/8} {\left[
\left(\partial_1 E\right) \left(\A_2\right)^2 + 
\left(\partial_2 E\right) \left(\A_1\right)^2 -
2\, \sqrt{\partial_1 E \partial_2 E - E \partial_1\partial_2 E}\; \A_1 \A_2
\right]}\over
{\left( \A_1 \B_2 + \A_2 \B_1 \right)^2}}\, ,
\nonumber\\
G_{33} & = &
{ E^{3/8} {\left[
\left(\partial_1 E\right) \left(\B_2\right)^2 + 
\left(\partial_2 E\right) \left(\B_1\right)^2 +
2\, \sqrt{\partial_1 E \partial_2 E - E \partial_1\partial_2 E}\; \B_1 \B_2
\right]}\over
{\left( \A_1 \B_2 + \A_2 \B_1 \right)^2}}\, .
\label{bw}
\end{eqnarray}
It is worth pointing out that such simplification was possible only 
because the coefficients $S_{11}$, $S_{12}$ and $S_{22}$ appear all 
in the form of a complete square which is a consequence of imposing 
the ans\"taz (\ref{au}). In other words, even if we have not imposed 
the ans\"atz (\ref{au}) at the beginning we would have been driven 
to it by insisting that $S_{11}$, $S_{13}$ and $S_{33}$ are each in 
a form of a complete square in order to obtain at the end a solvable 
linear system of equations.
\smallskip
 
The last simplification occuring in our problem uses 
the Bianchi identity on the R-R field strength of the
configuration and which effectively relates the functions 
$\A_1$ and $\A_2$ to each other and similarly for the 
functions $\B_1$ and $\B_2$. That is we have:
\begin{eqnarray}
\left({{\A_1}\over{E}}\right)^2 & = & \partial_1\, A_{t12}\,,\qquad
\left({{\A_2}\over{E}}\right)^2 = \partial_2\, A_{t12}\,,
\label{bp}\\
\left({{\B_1}\over{E}}\right)^2 & = & -\; \partial_1\, A_{t34}\, ,\qquad
\left({{\B_2}\over{E}}\right)^2 = -\; \partial_2\, A_{t34}\, ,
\label{bq}
\end{eqnarray}
where $A_{\mu_1 \mu_2 \mu_3}$ is the 3-form potential an
$\partial_{1,2} \equiv \partial\, /\partial h_{1,2}$. 
Therefore, the low-energy solution of D2-branes 
at angles is completely determined by the knowledge of
$\left(A_{t12} , A_{t34} , E\right)$.
\smallskip

We will proceed in the following with the resolution 
of the linear system of equations (\ref{bm},\ref{bn},\ref{bo}) 
to find the metric components $G_{11}$, $G_{13}$ and $G_{33}$.
For this, we need to assume an ans\"atz for each function
in the set $\left(A_{t12} , A_{t34} , E\right)$. By examining
the special supergravity solutions of Section~(3.1), Section~(3.2) 
and Section~(3.3), we arrive to these ans\"atze for the background fields 
$\left(A_{t12} , A_{t34} , E\right)$:
\begin{eqnarray}
A_{t12} & = &
{{p_{11} \, h_1 + p_{22} \, h_2 + p_{12} \, h_1 \, h_2 }\over E}\, ,
\label{br}\\
A_{t34} & = &
{{q_{11} \, h_1 + q_{22} \, h_2 + q_{12} \, h_1 \, h_2}\over E}\, ,
\label{bs}\\
E & = & l_{00} + l_{11} \, h_1 + l_{22} \, h_2 + l_{12} \, h_1 \, h_2 \, .
\label{bt}
\end{eqnarray}
To determine the coefficients $\left(q_{ab} , p_{ab} , l_{ab}\right)$,
$a,b = 1,2$, we refer back again to the special limit of two parallel 
D2-branes, two orthogonal D2-branes and a single rotated D2-brane
given in (\ref{ag},\ref{aj},\ref{am}), respectively. Then by a 
straightforward comparison with these special, we make the following
identifications:
\begin{eqnarray}
p_{11} & = & \cos^2 \theta\,,\qquad p_{22} = 1\, ,
\nonumber\\
q_{11} & = & -\, \sin^2 \theta\,,\qquad q_{22} = 0\, ,
\nonumber\\
l_{00} & = & l_{11}  =  l_{22} = 1\, ,
\nonumber\\
p_{12} & = & -\, q_{12} = l_{12} \equiv l(\theta)\, .
\label{bu}
\end{eqnarray}
After replacing by the coefficients above into equations 
(\ref{bp},\ref{bq},\ref{bt}), we obtain 
\begin{eqnarray}
\partial_1 E & = & 1 + l\left(\theta\right) h_2\, ,\quad
\partial_2 E = 1 + l\left(\theta\right) h_1\, , \quad
\partial_1 E \partial_2 E - E \partial_1\partial_2 E = 1 - 
l\left(\theta\right)\, ,
\nonumber\\
E & = & 1 + h_1 + h_2 + l\left(\theta\right)\, h_1 \, h_2\, ,
\nonumber\\ 
\left(\A_1\right)^2 & = &
\cos^2 \theta + h_2\, \left[l\left(\theta\right) - \sin^2 \theta \right]\, ,
\nonumber\\
\left(\A_2\right)^2 & = &
\left(1 + h_1\, \sin^2 \theta \right)\,
\left[ 1 + h_1 \, l\left(\theta\right) \right]\, ,
\nonumber\\
\left(\B_1\right)^2 & = &
\left( 1 + h_2 \right)\,
\left[ \sin^2 \theta + l\left(\theta\right)\, h_2 \right]\, ,
\nonumber\\
\left(\B_2\right)^2 & = &
\left( h_1\, \cos \theta \right)^2 \, l\left(\theta\right) +
h_1\, \left[ l\left(\theta\right) - \sin^2 \theta \right]\, ,
\nonumber\\
\left(\C_1\right)^2 & = & \A_1 \, B_1\;, \qquad 
\left(\C_2\right)^2 = \A_2 \, \B_2 \, .
\label{bv}
\end{eqnarray}
\smallskip

It is clear that we have not yet identified completely the
metric components $G_{11}$, $G_{13}$ and $G_{33}$ in (\ref{bw})
since they still depend on the unknown function $l\left(\theta\right)$ 
through the parameters $(\A_1 , \A_2 , \B_1 , \B_2 , E)$. To find
the function $l\left(\theta\right)$, we use the relations in (\ref{bv}) 
to substitute by $\left(U,V\right)$ of (\ref{bh}), 
$\left(S_{11} , S_{12} , S_{22}\right)$ of 
(\ref{be},\ref{bf},\ref{bg}) and by 
$\left(G_{11}, G_{13}, G_{33}\right)$
of (\ref{bw}) into the equation of motion of (\ref{bcc}). Requiring 
consistency of this equation of motion (after a considerable amount
of algebra) yields the simple answer
\begin{equation}
l\left(\theta\right) = \sin^2 \theta\; .
\label{bx}
\end{equation}
Our metric and background fields are thus finally given by
\begin{eqnarray}
E & = & 1 + h_1 + h_2 + \sin^2 \theta\; h_1 \, h_2 \, ,
\nonumber\\
&&
\nonumber\\
A & = & {{h_2 + \cos^2 \theta h_1 + \sin^2 \theta h_1 h_2}\over E}\, 
\d t \wedge \d x^1 \wedge \d x^2 - {{\sin^2 \theta \left(h_1 + h_1 
h_2\right)}\over E}\, 
\d t \wedge \d x^3 \wedge \d x^4
\nonumber\\
& & \;\; +\, {{\cos \theta \sin \theta h_1}\over E}\, 
\left(\d t \wedge \d x^2 \wedge \d x^3 + \d t \wedge \d x^2 \wedge \d x^4\right) 
\, ,
\nonumber\\
U^2 & = & E^{-\, 5/8}\, ,\qquad V^2 = E^{3/8}\, ,\qquad e^{2\phi} = \sqrt{E}\, ,
\nonumber\\ 
G_{13} & = & E^{3/8}\,
{{\cos \theta\, \sin \theta \, h_1}\over{{E}}}\, ,
\nonumber\\ 
G_{11} & = & E^{3/8}\,
{{1 + \sin^2 \theta \, h_1}\over{{E}}}\, ,
\nonumber\\
G_{33} & = & E^{3/8}\,
{{1 + h_2 + \cos^2 \theta \, h_1}\over{{E}}}\, .
\label{by}
\end{eqnarray}
This completes our description of the solution of the bound
state of two D2-branes at angles which is seen to match the 
one guessed (but not derived) by the authors of~[9]. The advantage
of our approach here is that we tried to rely as much as possible
only on the equations of motion and the knowledge of some well know
limits of the general bound state with angle. Of course the
configuration of D2-branes at angles is simple enough that
its background fields can be written down without really 
having to solve any field equations. For more complicated 
(and higer dimensional) D-brane bound states, however, 
a pure guess work would not be enough. In this respect,
he D2-brane bound state with angle treated here 
by solving directly the equation serves as a `theoretical
laboratory' to illustrate the kind of calculation that is
involved in the general case. Indeed, in the next section 
we will take this step further and apply what we have learned 
from dealing with the D2-brane example to the case of D4-branes 
at angles. Our conclusion will be that a similar calculation can 
be applied for any similar Dp-branes at angles.  
\smallskip

\section{Generalization to D4-Branes at Angles}

In this section, we show how the previous calculation can be adapted 
to account for the case of a bound state of two D4-branes 
intersecting at non-trivial angles. As for the case of D2-branes, 
we start by writing down the equations of motion of a D4-brane. 
They are given by
\begin{equation}
{R^\mu}_\nu = {1\over 2}\, \partial^\mu \phi\, \partial_\nu \phi +
{1\over{2\, 4!}}\, e^{-\, \phi/2}\,
\left[4\, F^{\mu\alpha_2\alpha_3\alpha_4}\,F_{\nu\alpha_2\alpha_3\alpha_4}\, -\,
{{3}\over 8}\, {G^\mu}_\nu \,  F^2\right]\, ,
\label{da}
\end{equation}
\begin{equation}
\Box\phi = {{1}\over{4\, 4!}}\, e^{-\, \phi/2}\, F^2\, ,
\label{db}
\end{equation}
\begin{equation}
\partial_{\mu_1}\, 
\left(\sqrt{-G}\, e^{-\, \phi/2}\, F^{\mu_1 \mu_2 \mu_3 \mu_4}
\right) = 0 \, ,
\label{dc}
\end{equation}
\begin{equation}
\partial_{[ \nu}\, F_{\mu_1 \mu_2 \mu_3 \mu_4 ]} = 0\, .
\label{dd}
\end{equation}
The last equation is the Bianchi identity $d F = 0$ and we note
that the D4-brane is {\it magnetically} charged.
\smallskip

One can repeat the same steps done before for the D2-brane case and begin
first by constructing the configuration of two D4-branes related by $SU(2)$
rotations. We take one of the D4-branes, which is
described by the harmonic function $h_1$, to be parallel to 
$\left(y^1 , y^2 , x^3 , x^4\right)-$directions.
The worldvolume of the other D4-brane, which is represented by the harmonic 
function $h_2$, is chosen to oriented along the 
$\left(x^3 , x^4 , x^6 , x^7\right)-$axes.
For later reference and to also facilitate the comparison of the bound
state of D4-branes at angles to well known solutions we 
define the $y^{1,2,5,6}$ rotated axes as follows: 
\begin{eqnarray}
y^1 & = & \cos \theta\; x^1 - \sin \theta \; x^5\; , \qquad
y^5  =  \sin \theta \; x^1 + \cos \theta \; x^5\, ,
\nonumber\\
y^2 & = &\cos \theta \; x^2 + \sin \theta \; x^6\; , \qquad
y^6  =  -\, \sin \theta \; x^2 + \cos \theta \; x^6 \; .
\label{dda}
\end{eqnarray}
\smallskip

In the next step we explore the bound state of D4-branes with angle
in the special limits of two parallel D4-brane, two orthogonal 
D4-branes intersecting over a $2-$brane, and finally that of a 
single `tilted' D4-brane. The knowledge of the background fields 
in these configurations will allow us to extract the suitable 
ans\"atze that we later substitute into the equations of motion 
(\ref{da}--\ref{dd}). We do not show here all the details (since 
they are similar to D2-brane case) but we content ourselves by 
giving only the final results which (in the Einstein-frame metric)
are given by:
\begin{eqnarray}
\d s^2 & = &  U^2 \left(-\, \d t^2 + \left(\d x^3\right)^2 + \left(\d 
x^4\right)^2\right)
+ G_{11}\, \left( \left(\d x^1\right)^2 +  \left(\d x^2\right)^2 \right)
\nonumber\\
& & + G_{55}\, \left( \left(\d x^5\right)^2 +  \left(\d x^6\right)^2 \right) 
+ 2\, G_{15}\, \left( \d x^1\, \d x^5 - \d x^2\, \d x^6 \right)
+ V^2\, \displaystyle{\sum_{i=7,8,9}\left(\d x^i\right)^2}\, ,
\nonumber\\
^{\star} F & = &  -\;  
\partial_i\, \left[{{h_2 + \cos^2 \theta h_1 + 
l\left(\theta\right) h_1 h_2}\over E}\right]\, 
\d t \wedge \d x^1 \wedge \d x^2 \wedge \d x^3 \wedge \d x^4 \wedge \d x^i 
\nonumber\\ 
& &\;\; +\; \partial_i\, 
\left[{{\sin^2 \theta h_1 + l\left(\theta\right) h_1 h_2}\over E}\right]\, 
\d t \wedge \d x^3 \wedge \d x^4 \wedge \d x^5 \wedge \d x^6 \wedge \d x^i
\nonumber\\
& & \;\; -\; \partial_i\, 
\left[{{\cos \theta \sin \theta h_1}\over E}\right]\, 
\d t \wedge \left(\d x^2 \wedge \d x^5 + \wedge \d x^1 \wedge \d 
x^6\right)\wedge 
\d x^3 \wedge \d x^4 \wedge \d x^i \, ,
\nonumber\\
e^{-\, 2\phi} & = & \sqrt{1 + h_1 + h_2 + l\left(\theta\right) \, h_1 h_2}\; .
\label{de}
\end{eqnarray}
As in the D2-brane case, we have 
\begin{eqnarray}
U & \equiv & U\left(h_1 , h_2 , \theta\right),\;\;
V \equiv V\left(h_1 , h_2 , \theta\right),\;\;
E \equiv E\left(h_1 , h_2 , \theta\right), 
\nonumber\\
G_{11} & \equiv & G_{11}\left(h_1 , h_2 , \theta \right),\;\;
G_{13} \equiv G_{13}\left(h_1 , h_2 , \theta \right),\;\;
G_{33} \equiv G_{33}\left(h_1 , h_2 , \theta \right),
\nonumber\\
\A_m & \equiv & \A_m\left(h_1 , h_2 , \theta\right),\;\;
\B_m \equiv \B_m\left(h_1 , h_2 , \theta\right), \;\;
\C_m \equiv \C_m\left(h_1 , h_2 , \theta\right),\;\; m = 1,2\; ,
\label{df}
\end{eqnarray}
and the harmonic functions become
${h_{1,2}} = {{c_{1,2} \, Q_{1,2}}/{\left| \vec{r} - \vec{r}_{1,2} \right|}}$,
with $r^2 = \displaystyle{\sum_{i=7}^{9}\, (x^i)^2}$.
\smallskip

In order to determine the set of functions 
$\left(U , V , G_{11} , G_{13} , G_{33} , l\left(\theta\right)\right)$, 
we replace by the above ans\"atze into the field equations of motion 
(\ref{da}--\ref{dd}). The equations of motion are found to take the 
simple form:
\begin{eqnarray}
\partial_i \partial_i \phi & = & {1\over 4}\, {1\over{4!}}\, V^2 \, 
e^{-\phi/2}\, F^2 = {4\over 3}\; \partial_i \partial_i \ln\, U = 
-\; \partial_i \partial_i  \ln \left( G_{11} G_{55} - G_{15}^2 \right)\; ,
\label{dg}
\end{eqnarray}
\begin{eqnarray}
& & \partial_i \ln U\,\partial_j \ln U  -  
{1\over 2} \left(\partial_i G^{11}\, \partial_j G_{11} + \partial_i G^{55}\, 
\partial_j G_{55}
+ 2 \partial_i G^{15}\, \partial_j G_{15}\right) +
\partial_i \ln V\,\partial_j \ln V
\nonumber\\
& &\;\; + \delta_{ij}\, \partial_k \partial_k \ln V = 
- {1\over 2} \partial_i \phi\, \partial_j \phi 
+\, {3\over{8}}\,  V^2 \, \delta_{ij}\; {1\over{4!}}\, e^{-\phi/2}\, F^2  
- {2}\; {1\over{4!}}\, e^{-\phi/2}\, \left(F^2\right)_{ij}\, .
\label{dh}
\end{eqnarray}
Equation (\ref{dg}) leads readily to the solutions:
\begin{equation}
U = E^{-\, 3/16}\; ,\qquad G_{11}\, G_{55} - G_{15}^2 = E^{1/4}\; .
\label{di}
\end{equation}
Next to determine $V$ we need to use the extremality condition 
which follows from the requirement that the bound 
state of angled D4-branes preserves 1/4 of supersymmetries. 
An analysis similar to the D2-brane example leads in the present
case to this constraint
\begin{equation}
U^3\; \left( G_{11}\, G_{55} - G_{15}^2 \right)\; V = 1\, ,
\label{dj}
\end{equation}
which readily yields 
\begin{equation}
 V = E^{5/16}\, .
\label{dk}
\end{equation}
\smallskip

To continue with the identification of the rest of the background fields
of the angled D4-branes, we need to calculate the quantity 
$F^2$ from the ans\"atz field strength of (\ref{de}). We skip here all the 
details involved in the evaluation of $F^2$ but it is clear that at the end 
it will turn out to be a function of $\left(h_1 , h_2 , 
l\left(\theta\right)\right)$. 
Substituting then by its expression into the two equations of motion 
(\ref{dg},\ref{dh}), (and without giving any details since 
they are identical to those explained in Section~(4)),
we are lead to the rest of the supergravity solution which is 
found to be:
\begin{eqnarray}
l\left(\theta\right) & = &  \sin^2 \theta\; , \qquad
E = 1 + h_1 + h_2 + \sin^2 \theta \, h_1 h_2\, ,
\nonumber\\
e^{-\, 2 \phi} & = & \sqrt{E}\;\; , \quad U^2 = E^{-\, 3/8}\;\; , \quad
V^2 = E^{5/8}\; ,
\nonumber\\
& &
\nonumber\\
G_{15} & = & E^{-\, 3/8}\,
{{\cos \theta\, \sin \theta \, h_1}\over{{E}}}\, ,
\nonumber\\ 
G_{11} & = & E^{-\, 3/8}\,
{{1 + \sin^2 \theta \, h_1}\over{{E}}}\, ,
\nonumber\\
G_{55} & = & E^{-\, 3/8}\,
{{1 + h_2 + \cos^2 \theta \, h_1}\over{{E}}}\, ,
\nonumber\\
^{\star} F & = &  -  
\partial_i\, \left[{{h_2 + \cos^2 \theta h_1 + 
\sin^2 \theta h_1 h_2}\over E}\right]\, 
\d t \wedge \d x^1 \wedge \d x^2 \wedge \d x^3 \wedge \d x^4 \wedge \d x^i 
\nonumber\\ 
& & +\; \partial_i\, 
\left[{{\sin^2 \theta \, \left(h_1 + h_1 h_2\right)}\over E}\right]\, 
\d t \wedge \d x^3 \wedge \d x^4 \wedge \d x^5 \wedge \d x^6 \wedge \d x^i
\nonumber\\
& &  - \partial_i 
\left[{{\cos \theta \sin \theta h_1}\over E}\right] 
\d t \wedge \left(\d x^2 \wedge \d x^5 + \wedge \d x^1 \wedge \d 
x^6\right)\wedge 
\d x^3 \wedge \d x^4 \wedge \d x^i\, .
\label{dl}
\end{eqnarray}
So with the above background fields our 1/4 supersymmetric bound state of angled 
two D4-branes is totally defined. The supergravity solution of this 
configuration
was also conidered in~[8,9].
\smallskip

\section{Conclusions}

In this paper, we considered the supergravity solutions
describing D2-brane and D4-brane bound states where the 
constituent D-branes are oriented at non-trivial angles 
with respect to one another. Since the studied bound 
states are taken to saturate a BPS bound where 1/4 of 
supersymmetries are preserved, the constituent D-branes 
must be related by $SU(2)$ rotations. The originality of 
our work here resides in our use of the equations of motion 
to solve directly for the background fields of the angled 
D-branes. In fact, using always only the equations of motion, 
the calculation of this paper can be applied easily to derive 
the low-energy solution of any configuration of multiple Dp-branes 
oriented at angles. Finally, the work in this paper will serve as 
a basis to derive the supregravity classical solutions of more
general bound states with angles in type II string theory 
and M-theory~[18].
\smallskip

\section*{Acknowledgments}
This research is supported by the Institut des Hautes Etudes 
Scientifiques. The author is grateful to all the personnel
of the Institut des Hautes Etudes Scientifiques and especially
to Louis Michel and Thibault Damour for their support and 
hospitality. It is a pleasure also to acknowledge useful 
conversations with Thibault Damour, Costas Bachas, Dimitri Polyakov, 
R. C. Myers and Ramzi Khuri. Finally, we would like to thank
all the participants in the working groups at ENS and LPTHE 
(Jussieu) in Paris.

\end{document}